\documentclass[prd,aps,tightline,twocolumn,nofootinbib,superscriptaddress]{revtex4}

\usepackage{amssymb}
\usepackage{amsmath}
\usepackage{graphicx}
\usepackage{latexsym}

\begin{document}

\begin{flushleft}
STUPP-09-205,  TU-861, YITP-10-1, ICRR-Report-556
\end{flushleft}


\title{ %
  Stau relic density at the Big-Bang nucleosynthesis era consistent with the
  abundance of the light element nuclei in the coannihilation
  scenario
}

\author{Toshifumi Jittoh}
\affiliation{Department of Physics, Saitama University, 
     Shimo-okubo, Sakura-ku, Saitama, 338-8570, Japan}

\author{Kazunori Kohri}
\affiliation{Physics Department, Lancaster University LA1 4YB, UK}
\affiliation{Department of Physics, Tohoku University, Sendai 980-8578, Japan}

\author{Masafumi Koike}
\affiliation{Department of Physics, Saitama University, 
     Shimo-okubo, Sakura-ku, Saitama, 338-8570, Japan}

\author{Joe Sato}
\affiliation{Department of Physics, Saitama University, 
     Shimo-okubo, Sakura-ku, Saitama, 338-8570, Japan}
     
\author{Takashi Shimomura}
\affiliation{Departament de $F\acute isica$ $Te\grave orica$ and IFIC, 
Universitat de $Val\grave encia$-CSIC, E-46100 Burjassot, $Val\grave encia$, Spain}
\affiliation{Yukawa Institute for Theoretical Physics, Kyoto University, Kyoto 606-8502, Japan}     

\author{Masato Yamanaka}
\affiliation{Institute for Cosmic Ray Research, University of Tokyo, Kashiwa 277-8582, Japan}


\begin{abstract}
We calculate the relic density of stau at the beginning of the Big-Bang 
Nucleosynthesis (BBN) era in the coannihilation scenario of minimal 
supersymmetric standard model (MSSM). In this scenario, stau 
can be long-lived and form bound states with nuclei. 
We put constraints on the parameter space of MSSM by connecting the calculation of the
relic density of stau to the observation of the light elements abundance, which
strongly depends on the relic density of stau.
Consistency between the theoretical prediction and the observational
result, both of the dark matter abundance and the light elements abundance,
requires the mass difference between the lighter stau and the lightest neutralino to be around 100MeV,
the stau mass to be 300 -- 400 GeV, and the mixing angle of the left and right-handed staus to be 
$\sin\theta_{\tau} = (0.65 \textrm{ -- } 1)$. 
\end{abstract}


\maketitle

\section{Introduction}    

Cosmological observations have established the existence of the non-baryonic 
dark matter (DM) \cite{Dunkley:2008ie}. These observations suggest that 
the DM is a stable and weakly-interacting particle with a mass of 
$\mathcal{O}(100)$ GeV. Many hypothetical candidates for the DM have 
been proposed in models of particle physics beyond the standard model (SM), 
and one of the most attractive candidates is the lightest neutralino, 
$\tilde{\chi}^0$, in supersymmetric extensions of the SM with $R$ parity 
conservation. Neutralino is a linear combination of the superpartners of $U(1),~SU(2)$ 
gauge bosons and the two neutral Higgses, and is stable when it is the lightest 
supersymmetric particle (LSP). Indeed it accounts for the observed DM 
abundance when it is degenerate in mass to the next lightest supersymmetric 
particle (NLSP) and hence coannihilates with the NLSP \cite{Griest:1990kh}. 
We consider the setup that the LSP is a neutralino consisting of mainly bino, 
the superpartner of $U(1)$ gauge boson, and the NLSP is the lighter stau, the 
superpartner of tau lepton. 
This is naturally realized in the 
MSSM with the unification condition at the grand unified theory scale.  The 
minimal supersymmetric SM (MSSM) has two eigenstates of stau as physical state.
In 
absence of inter-generational mixing the mass eigenstate of stau is given 
by the linear combination of the left-handed stau $\tilde \tau_L$ and the 
right-handed stau $\tilde \tau_R$ as
\begin{equation}
 \begin{split}
   \tilde \tau = \cos \theta_\tau  \tilde \tau_L  
   +  \sin  \theta_\tau  \text{e}^{-i \gamma_\tau}  \tilde \tau_R ,
 \end{split}
\end{equation}
where $\theta_\tau$ is the left-right mixing angle and $\gamma_\tau$ is the 
CP violating phase.

In a scenario of the coannihilation, the NLSP stau can be long-lived if the mass 
difference, $\delta m$, between neutralino and stau is  small enough 
to forbid two-body decays of stau into neutralino. It was shown in 
\cite{Jittoh:2005pq} that the lifetime of stau is longer than $1000$ second 
for $\delta m \lesssim 100$ MeV. It is known \cite{Cyburt:2009pg, 
Kamimura:2008fx, Kusakabe:2008kf, Pospelov:2008ta, Kaplinghat:2006qr, 
Jedamzik:2007qk, Kawasaki:2008qe, Pradler:2007is, Jedamzik:2007cp, Jittoh:2007fr, Jittoh:2008eq, 
Bird:2007ge, Kohri:2006cn, Pospelov:2006sc, Bailly:2008yy} that the long-lived charged particles 
affect the relic abundance of the light nuclei during or after the big-bang 
nucleosynthesis (BBN).

There is a  discrepancy, the so-called $^7$Li problem, on the
primordial $^7$Li abundance between the prediction from the standard BBN (SBBN)
and the observations \cite{Cyburt:2008kw, Spite:1982dd}. Combined with the
up-to-date values of baryon-to-photon ratio, $\eta=(6.225\pm 0.170) \times
10^{-10}$ from Wilkinson Microwave Anisotropy Probe (WMAP)
\cite{Dunkley:2008ie}, the SBBN predicts the $^7\mathrm{Li}$ to proton ratio,
$(^7\mathrm{Li/H})_\mathrm{SBBN}=5.24^{+0.71}_{-0.67} \times 10^{-10}$, which is
by about four-times larger than its observed value in poor-metal halos
\cite{Cyburt:2008kw, Cyburt:2008up}. Because there exists no general agreements
about astrophysical scenarios to reduce the $^7$Li abundance
{\cite{Richard:2004pj,Korn:2006tv,Melendez:2009xv,Lind:2009ta}}, it is natural
to consider  nonstandard effects.

The authors have investigated the BBN including the long-lived 
stau \cite{Jittoh:2007fr, Jittoh:2008eq}. The long-lived stau form a bound 
state with nuclei $(\tilde{\tau}N)$, and consequently convert it into a 
nucleus with a smaller atomic number. Here $N$ stands for a nucleus. 
In this
scenario, the abundance of $^7$Li is reduced through the conversion process,
$(\tilde{\tau} {}^7\mathrm{Be}) \rightarrow \tilde{\chi}^0 + \nu_\tau +
^7\!\mathrm{Li}$ and the further destruction of $^7\mathrm{Li}$ by either
a collision with a background proton or another conversion process,
 $(\tilde{\tau} {}^7\mathrm{Li}) \rightarrow
\tilde{\chi}^0 + \nu_\tau + ^7\!\mathrm{He}$.  Therefore, the more these 
bound states are formed, the more the $^7$Li abundance is reduced. The 
number density of the bound state is determined by the relic density of stau. In
\cite{Jittoh:2007fr, Jittoh:2008eq}, we assumed 
$Y_{\tilde{\tau}, \mathrm{FO}}$ and $\delta m$ to be free parameters of 
the scenario, where $Y_{\tilde{\tau}, \mathrm{FO}}$ is the yield value of 
stau at the time of decoupling from the thermal bath. The full calculation of the light 
nucleus abundances showed a region in $(\delta m, Y_{\tilde{\tau},
\mathrm{FO}})$ plane where the $^7$Li problem is solved consistently with the
observational constraints on the other nuclei. The region points $Y_{\tilde{\tau},
\mathrm{FO}}$ to be close to the yeild value of the DM and $\delta m$ to be 
around $0.1$ GeV, respectively.

In this work, we improve our previous analysis by calculating 
$Y_{\tilde{\tau},\mathrm{FO}}$ with taking the relic abundance of 
DM into account. The outline of this paper is as follows. In section \ref{cal}, we 
review the formalism for the calculation of the relic density of stau, and 
derive the Boltzmann equations of stau and neutralino for the 
calculation. In section \ref{numerical}, we present the numerical results of the 
calculation, and prove the parameter space for solving the $^7$Li problem. 
Section \ref{summary} is devoted to a summary and discussion.

\section{formalism for the calculation of relic density of stau at the BBN era}  \label{cal} 

In this section, we prepare to calculate the relic density of stau at the BBN 
era. Firstly, in subsection \ref{bol}, we briefly review the
Boltzmann equations for the number density of stau and neutralino based
on the thermal relic scenario. Then, in subsection \ref{evo}, we discuss 
the number density evolution of stau and neutralino quantitatively. In 
subsection \ref{ex pro}, we investigate the significant processes for the
calculation of the relic density of stau. Finally, we obtain the Boltzmann
equations for the relic density of stau in a convenient form.

\subsection{Boltzmann equations for the number density evolution of stau and neutralino}    \label{bol} 

In this subsection, we show the Boltzmann equations of stau and
neutralino and briefly review their quantitative structure based on
\cite{Edsjo:1997bg} (see also a recent paper\cite{Berger:2008ti}).

We are interested in the relic density of stau in the coannihilation scenario. In this 
scenario, stau and neutralino are quasi-degenerate in mass and decouple 
from the thermal bath almost at the same time \cite{Griest:1990kh}. Thus the 
relic density of stau is given by solving a coupled set of the Boltzmann equations 
for stau and neutralino as simultaneous differential equation. 
For simplicity, we use the Maxwell-Boltzmann statistics for all species instead of the
Fermi-Dirac for fermions and the Bose-Einstein for bosons, and assume T invariance. 
With these simplifications, the Boltzmann equations of them are given as follows
\begin{equation}
 \begin{split}
   &\frac{dn_{\tilde \tau^-}}{dt} + 3H n_{\tilde \tau^-}  = \\
   &- \sum_{i } \sum_{X, Y} \langle \sigma v \rangle_{\tilde \tau^- i 
   \leftrightarrow X Y}  
   \biggl[ n_{\tilde \tau^-} n_i - n_{\tilde \tau^-}^{eq} n_i^{eq} 
   \left( \frac{n_X n_Y}{n_X^{eq} n_Y^{eq}} \right) \biggr]   \\
   &- \sum_{i \neq \tilde \tau^-} \sum_{X, Y} 
   \Biggl\{   \langle \sigma' v \rangle_{\tilde \tau^- X \rightarrow i Y} 
   \biggl[ n_{\tilde \tau^-} n_X \biggr]  
   - \langle \sigma' v \rangle_{i Y \rightarrow \tilde \tau^- X} 
   \biggl[ n_{i} n_Y \biggr] \Biggr\}
 \end{split}     \label{a}
\end{equation}
\begin{equation}
 \begin{split}
   &\frac{dn_{\tilde \tau^+}}{dt} + 3H n_{\tilde \tau^+}  = \\
   &- \sum_{i } \sum_{X, Y} \langle \sigma v \rangle_{\tilde \tau^+ i 
   \leftrightarrow X Y} 
   \biggl[ n_{\tilde \tau^+} n_i - n_{\tilde \tau^+}^{eq} n_i^{eq} 
   \left( \frac{n_X n_Y}{n_X^{eq} n_Y^{eq}} \right)  \biggr]   \\
   &- \sum_{i \neq \tilde \tau^+} \sum_{X, Y} 
   \Biggl\{   \langle \sigma' v \rangle_{\tilde \tau^+ X \rightarrow i Y} 
   \biggl[ n_{\tilde \tau^+} n_X \biggr] 
   - \langle \sigma' v \rangle_{i Y \rightarrow \tilde \tau^+ X} 
   \biggl[ n_{i} n_Y \biggr] \Biggr\}
 \end{split}     \label{b}
\end{equation}
\begin{equation}
 \begin{split}
   &\frac{dn_{\tilde \chi}}{dt} + 3H n_{\tilde \chi}  =   \\
   &- \sum_{i } \sum_{X, Y} \langle \sigma v \rangle_{\tilde \chi i 
   \leftrightarrow X Y} 
   \biggl[ n_{\tilde \chi} n_i - n_{\tilde \chi}^{eq} n_i^{eq} 
   \left( \frac{n_X n_Y}{n_X^{eq} n_Y^{eq}} \right)   \biggr]   \\
   &- \sum_{i \neq \tilde \chi} \sum_{X, Y} 
   \Biggl\{   \langle \sigma' v \rangle_{\tilde \chi X \rightarrow i Y} 
   \biggl[ n_{\tilde \chi} n_X \biggr] 
   - \langle \sigma' v \rangle_{i Y \rightarrow \tilde \chi X} 
   \biggl[ n_{i} n_Y \biggr] \Biggr\} ~.
 \end{split}    \label{c}
\end{equation}
Here $n$ and $n^{eq}$ represent the actual number density and the equilibrium number density of
each particle, and $H$ is the Hubble expansion rate. Index $i$ denotes stau
and neutralino, and indices $X$ and $Y$ denote SM particles. Note that if
relevant SM particles are in thermal equilibrium, $n_X = n_X^{eq}$, $n_Y =
n_Y^{eq}$, and $( n_X n_Y / n_X^{eq} n_Y^{eq} ) = 1$ then these equations are
reduced into a familiar form. $\langle \sigma v \rangle$ and $\langle \sigma' v
\rangle$ are the thermal averaged cross sections, which is defined by
\begin{equation}
 \begin{split}
   \langle \sigma v \rangle_{12 \rightarrow 34} &\equiv 
   g_{12}\frac{\int d^3 {\bf p_1} d^3 {\bf p_2} ~ f_1 f_2 ~ 
   (\sigma v)_{12 \rightarrow 34}}
  {\int d^3 {\bf p_1} d^3 {\bf p_2} ~ f_1 f_2 }    \\
  &= g_{12}\frac{\int d^3 {\bf p_1} d^3 {\bf p_2} ~ f_1 f_2 ~ 
   (\sigma v)_{12 \rightarrow 34}}
  { n_1^{eq} n_2^{eq} }, 
 \end{split}     
\end{equation}
where $f$ is the distribution function of a particle, $v$ is the
relative velocity between initial state particles, and $g_{12} =2 (1)$
for same (different) particles 1 and 2. In this work, we assume that all
of the supersymmetric particles except for stau and neutralino
are heavy, and therefore do not involve them in the coannihilation processes.

The first line on the right-hand side of Eqs. (\ref{a}), (\ref{b}), and (\ref{c}) 
accounts for the annihilation and the inverse annihilation processes of the supersymmetric particles 
($i j \leftrightarrow  X Y$). Here index $j$ denotes stau and neutralino. As long 
as the R-parity is conserved, as shown later, the final number density of neutralino DM 
is controlled only by these processes. 
The second line accounts for the exchange processes by scattering off the cosmic 
thermal background ($i X  \leftrightarrow j Y$). These processes exchange stau with 
neutralino and vice versa, and thermalize them. Consequently, the number 
density ratio between them is controlled by these processes. Instead, these processes 
leave the total number density of the supersymmetric particles.
Note that in general, although there are terms which account for decay and inverse 
decay processes of stau ($\tilde \tau \leftrightarrow \tilde \chi X Y ...$) in the 
Boltzmann equations, we omit them. It is because we are interested in solving the 
$^7$Li problem by the long-lived stau, and the whole intention of this work is to search 
parameters which can provide the solution for the $^7$Li problem.  Hence we assume 
that the stau is stable enough to survire until the BBN era, and focusing on the mass difference 
between stau and neutralino is small enough to make it possible.

\subsection{The evolution of the number density of stau and neutralino}  \label{evo}    %

In this subsection, we discuss the evolution of the number density of each species. 
Firstly, we discuss the number density evolution of neutralino DM. Since we 
have assumed R-parity conservation, all of the supersymmetric particles eventually 
decay into the LSP neutralino. Thus its final number density is simply described by 
the sum of the number density of all the supersymmetric particles :
\begin{equation}
 \begin{split}
   N = \sum_i n_i ~.
 \end{split}     \label{d}
\end{equation}
For $N$, that is the number density of the neutralino, we get the Boltzmann 
equation by summing up Eqs. (\ref{a}), (\ref{b}), and (\ref{c}), 
\begin{equation}
 \begin{split}
   \frac{dN}{dt} + 3H N  = 
   - \langle \sigma v \rangle_{sum} \biggl[ N N - N^{eq} N^{eq} \biggr]
 \end{split}     \label{e}
\end{equation}
\begin{equation}
 \begin{split}
   \langle \sigma v \rangle_{sum} \equiv \sum_{i = \tilde \chi, \tilde \tau} \sum_{X,Y} 
   \langle \sigma v \rangle_{\tilde \chi i \leftrightarrow X Y} ~.
 \end{split}     \label{f}
\end{equation}
Notice that the terms describing the exchange processes in each
Boltzmann equations cancel each other out. Solving the
Eq. (\ref{e}), we obtain $N$ and find the freeze out temperature of the
total number density of all the supersymmetric particles $T_f$ by using the
standard technique \cite{Edsjo:1997bg}:
\begin{equation}
 \begin{split}
   \frac{m_{\tilde \chi}}{T_f} = 
   \text{ln} \frac{0.038 ~ g ~ m_{pl} ~m_{\tilde \chi} \langle \sigma v \rangle}
   {g_*^{1/2} (m_{\tilde \chi} / T_f)} \simeq 25  ~.
 \end{split}     \label{g}
\end{equation}
Here, $g$ and $m_{\tilde \chi}$ are the internal degrees of freedom and the mass of 
neutralino, respectively. The Planck mass $m_{pl} = 1.22 \times 10^{19}$ GeV, and $g_*$ 
are the total number of the relativistic degrees of freedom. Consequently, we see that 4 
GeV $\lesssim T_f  \lesssim$ 40 GeV for 100 GeV $\lesssim m_{\tilde \chi} \lesssim$ 
1000 GeV.

Now we will discuss the number density evolution of stau. To obtain the relic  
density of stau, we solve a coupled set of the Boltzmann equations, (\ref{a}), 
(\ref{b}), and (\ref{c}), as simultaneous differential equation. Each Boltzmann 
equation contains the contributions of the exchange processes. These processes 
exchange stau with neutralino and vice versa. At the 
temperature $T_f$, the interaction rate of the exchange processes is much larger 
than that of the annihilation and the inverse annihilation processes. This is because 
the cross sections of the exchange processes are in the same order of magnitude as 
that of the annihilation and the inverse annihilation, but the number density of the SM 
particles is much larger than that of the supersymmetric particles which is suppressed 
by the Boltzmann factor. Thus even if the total number density of stau and 
neutralino is frozen out at the temperature $T_f$, each number density of them 
continue to evolve through the exchange processes.

Thus, to calculate the relic density of stau, we have to follow the two-step 
procedures. As a first step, we calculate the total relic density of the supersymmetric 
particles by solving the Eq. (\ref{e}). We use the publicly available program 
micrOMEGAs \cite{Belanger:2008sj} to calculate it. The second step is the calculation 
of the number density ratio of stau and neutralino. The second step is significant 
for calculating the relic density of stau at the BBN era, and hence we will discuss it in 
detail in the next subsection.

\subsection{The exchange processes and Lagrangian for describing them}   \label{ex pro}    

After the freeze-out of the total number density of stau and neutralino, each of them is 
exchanged through the following processes
\begin{equation}
 \begin{split}
   \widetilde \tau \gamma &~\longleftrightarrow~ \widetilde \chi \tau \\
   \widetilde \chi \gamma ~ &~\longleftrightarrow~ \widetilde \tau \tau.
 \end{split}     \label{exchange}
\end{equation}
Notice that although there are other exchange processes via weak interaction (for 
example, $\widetilde \tau W \leftrightarrow \widetilde \chi \nu_\tau$, 
$\widetilde \tau \nu_\tau \leftrightarrow \widetilde \chi W$, and so on), we can omit 
them. This is because the number density of W boson is not enough to work these 
processes sufficiently due to the Boltzmann factor suppression, and the final state W boson is 
kinematically forbidden when the thermal bath temperature is less than $T_f$.  These 
processes (Eq. (\ref{exchange})) are described by the Lagrangian 
\begin{equation}
 \begin{split}
   \mathcal{L} ~&=~ 
    \tilde \tau^\ast  \overline{ \tilde \chi^0 }  ( g_L P_L + g_R P_R )  \tau   \\
    &- i e ( \tilde \tau^\ast (\partial_\mu \tilde \tau ) - (\partial_\mu \tilde \tau^\ast) \tilde \tau ) A^\mu
    +  \text{h.c.}, 
 \end{split}       \label{Lag}
\end{equation}
where $e$ is the electromagnetic coupling constant, $P_L$ and $P_R$ are
the projection operators, and $l \in \{ e, \mu \}$. $g_L$ and $g_R$ are the
coupling constants, given by
\begin{equation}
 \begin{split}
   g_L &= \frac{g}{\sqrt{2} \cos \theta_W}  \sin \theta_W  \cos \theta_\tau ,   \\
   g_R &= \frac{\sqrt{2} g}{ \cos \theta_W}  \sin \theta_W  \sin \theta_\tau \mathrm{e}^{i \gamma_\tau }, 
 \end{split}
\end{equation}
where $g$ is the $SU(2)_L$ gauge coupling constant, and $\theta_W$ is the Weinberg 
angle.

The evolution of the stau number density is governed only by the exchange processes 
(Eq.(\ref{exchange})) after the freeze-out of the total relic density of stau and 
neutralino. When we calculate it, we should pay attention to two essential points relevant 
to the exchange processes.

One is the competition between the interaction rate of the exchange processes and the 
Hubble expansion rate, since when these interaction rates get smaller than the Hubble 
expansion rate, the relic density of stau would be frozen out. 
The other is whether tau leptons are in the thermal bath or not. The interaction rate of 
the exchange processes strongly depends on the number density of tau leptons. When 
tau leptons are in the thermal bath, the number density ratio between stau and neutralino 
are given by the thermal ratio,
\begin{equation}
\label{ratio}
\frac{n_{\tilde{\tau}}}{n_\chi}
\simeq \frac{e^{-m_{\tilde{\tau}/T}}}{e^{-m_{\tilde \chi/T}}}
= \exp \Bigl(-\frac{\delta m}{T}\Bigr) ,
\end{equation}
through the exchange processes.  On the contrary, once tau leptons decouple from the 
thermal bath, the ratio cannot reach this value. To calculate the relic density of stau, we 
have to comprehend the temperature of tau lepton decoupling.

\begin{figure} [t!]
\begin{center}
\includegraphics[width=230pt,clip]{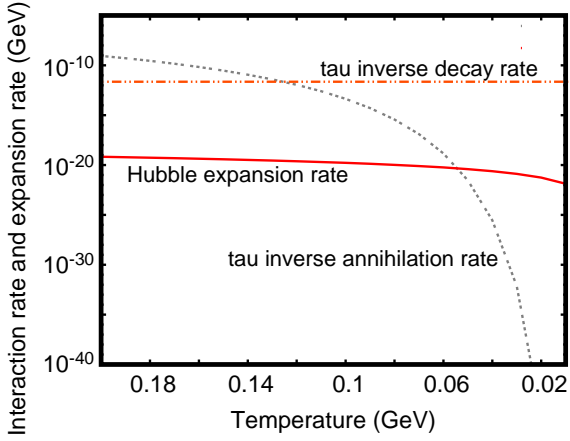}
\caption{{\small{The inverse annihilation rate of tau leptons $\langle \Gamma 
\rangle n_X^{eq}$, the inverse decay rate of tau leptons $\langle \Gamma \rangle$, 
and the Hubble expansion rate $H$ as a function of the thermal bath temperature. }} }
\label{tau}
\end{center}
\end{figure}

To see whether tau leptons are in the thermal bath or not, we consider the Boltzmann 
equation for its number density, $n_{\tau}$, 
\begin{equation}
 \begin{split}
   \frac{dn_{\tau}}{dt} + 3H n_{\tau} =  
   &- \langle \sigma v \rangle 
   \Biggl[  n_{\tau} n_{X} - n_{\tau}^{eq} n_{X}^{eq} 
   \left( \frac{n_Y n_Z}{n_Y^{eq} n_Z^{eq}} \right)  \Biggr]   \\
   &- \langle \Gamma \rangle \Biggl[  n_{\tau} - n_{\tau}^{eq} 
   \left( \frac{n_X n_Y ...}{n_X^{eq} n_Y^{eq} ...} \right) \Biggr] ,
 \end{split}
\end{equation} 
\begin{equation}
 \begin{split}
   \langle \sigma v \rangle = \sum_{X, Y, Z} 
   \langle \sigma v \rangle_{\tau X \leftrightarrow Y Z} ~, ~~~
   \langle \Gamma \rangle = \sum_{X, Y, ...} 
   \langle \Gamma \rangle_{\tau \leftrightarrow X Y ...}, 
 \end{split}
\end{equation} 
where indices $X$, $Y$, and $Z$ denote the SM particles, and $\langle \Gamma \rangle$ represents 
the thermal averaged decay rate of tau lepton. When the SM particles $X$, $Y$, and $Z$ are in the 
thermal equilibrium, $(n_Y n_Z)/(n_Y^{eq} n_Z^{eq}) = (n_X n_Y ...)/(n_X^{eq} n_Y^{eq} 
...) = 1$, and hence tau leptons are sufficiently produced through the inverse annihilation and/or 
the inverse decay processes as long as these interaction rates are larger than the Hubble 
expansion rate. Therefore, whether tau leptons are in the thermal bath or not can be distinguished  
by comparing the Hubble expansion rate $H$ with the inverse annihilation rate of tau lepton 
$ \langle \sigma v \rangle n_{X}^{eq}$, and the inverse decay rate of tau lepton $\langle 
\Gamma \rangle$.  In other words, the inequality expression
\begin{equation}
 \begin{split}
    \langle \sigma v \rangle n_X^{eq}  ~>~ H   
    ~~~~~\text{and/or}~~~~~  \langle \Gamma \rangle  ~>~  H
 \end{split}     \label{compare}
\end{equation}
indicates that tau leptons are in the thermal bath. Fig. 1 shows $\langle \Gamma \rangle 
n_X^{eq}$,  $\langle \Gamma \rangle$, and $H$ as a function of the thermal bath temperature. 
As shown in Fig. 1, the inverse decay rate of tau lepton is much larger than the Hubble expansion 
rate. Thus, we can conclude that tau leptons remain in the thermal bath still at the beginning of the 
BBN.

\subsection{Calculation of the number density ratio of stau and neutralino}    

We are now in a position to calculate the number density ratio of stau and neutralino. 
In this subsection, we will show a set of relevant Boltzmann equations.

The right-hand side of the Boltzmann equations (Eqs. (\ref{a}),
(\ref{b}), and (\ref{c})) depends only on temperature, and hence it is
convenient to use temperature $T$ instead of time $t$ as independent
variable. To do this, we reformulate the Boltzmann equations by using
the ratio of the number density to the entropy density $s$ :
\begin{equation}
 \begin{split}
   Y_i = \frac{n_i}{s} .
 \end{split}
\end{equation}
Consequently, we obtain the Boltzmann equations for the number density evolution 
of stau and neutralino
\begin{equation}
 \begin{split}
    &\frac{d Y_{\tilde \tau^-}}{dT} = \Bigl[ 3H T g_*(T) \Bigr]^{-1} 
      \biggl[ 3g_*(T) + T \frac{dg_*(T)}{dT} \biggr] s  \\
    &\times \Biggl\{ \langle \sigma v \rangle_{\tilde \tau^- \gamma 
      \rightarrow \tilde \chi \tau^-} Y_{\tilde \tau^-} Y_\gamma 
    - \langle \sigma v \rangle_{\tilde \chi \tau^- 
       \rightarrow \tilde \tau^- \gamma} Y_{\tilde \chi} Y_{\tau^-}  \\
    &+ \langle \sigma v \rangle_{\tilde \tau^- \tau^+ 
       \rightarrow \tilde \chi \gamma} Y_{\tilde \tau^-} Y_{\tau^+}     
    - \langle \sigma v \rangle_{\tilde \chi \gamma 
       \rightarrow \tilde \tau^- \tau^+} Y_{\tilde \chi} Y_{\gamma}       \Biggr\}
 \end{split}     \label{h}
\end{equation}
\begin{equation}
 \begin{split}
    &\frac{d Y_{\tilde \tau^+}}{dT} = \Bigl[ 3H T g_*(T) \Bigr]^{-1} 
      \biggl[ 3g_*(T) + T \frac{dg_*(T)}{dT} \biggr] s  \\
    &\times \Biggl\{ \langle \sigma v \rangle_{\tilde \tau^+ \gamma 
      \rightarrow \tilde \chi \tau^+} Y_{\tilde \tau^+} Y_\gamma 
    - \langle \sigma v \rangle_{\tilde \chi \tau^+ 
       \rightarrow \tilde \tau^+ \gamma} Y_{\tilde \chi} Y_{\tau^+}    \\
    &+ \langle \sigma v \rangle_{\tilde \tau^+ \tau^- 
       \rightarrow \tilde \chi \gamma} Y_{\tilde \tau^+} Y_{\tau^-}     
    - \langle \sigma v \rangle_{\tilde \chi \gamma 
       \rightarrow \tilde \tau^+ \tau^-} Y_{\tilde \chi} Y_{\gamma}       \Biggr\}
 \end{split}     \label{i}
\end{equation}
\begin{equation}
 \begin{split}
    &\frac{d Y_{\tilde \chi}}{dT} = \Bigl[ 3H T g_*(T) \Bigr]^{-1} 
      \biggl[ 3g_*(T) + T \frac{dg_*(T)}{dT} \biggr] s  \\
    &\times \Biggl\{ \langle \sigma v \rangle_{\tilde \chi \tau^- 
       \rightarrow \tilde \tau^- \gamma} Y_{\tilde \chi} Y_{\tau^-}   
    - \langle \sigma v \rangle_{\tilde \tau^- \gamma 
      \rightarrow \tilde \chi \tau^-} Y_{\tilde \tau^-} Y_\gamma    \\
    &+ \langle \sigma v \rangle_{\tilde \chi \gamma 
       \rightarrow \tilde \tau^- \tau^+} Y_{\tilde \chi} Y_{\gamma}  
    - \langle \sigma v \rangle_{\tilde \tau^- \tau^+ 
      \rightarrow \tilde \chi \gamma} Y_{\tilde \tau^-} Y_{\tau^+}  \\
    &+ \langle \sigma v \rangle_{\tilde \chi \tau^+ 
      \rightarrow \tilde \tau^+ \gamma} Y_{\tilde \chi} Y_{\tau^+}   
    - \langle \sigma v \rangle_{\tilde \tau^+ \gamma 
       \rightarrow \tilde \chi \tau^+} Y_{\tilde \tau^+} Y_\gamma   \\
    &+ \langle \sigma v \rangle_{\tilde \chi \gamma 
       \rightarrow \tilde \tau^+  \tau^-} Y_{\tilde \chi} Y_{\gamma}  
    - \langle \sigma v \rangle_{\tilde \tau^+  \tau^- 
       \rightarrow \tilde \chi \gamma} Y_{\tilde \tau^+} Y_{\tau^-}   \Biggr\} ~.
 \end{split}     \label{j}
\end{equation}
Here $g_*(T)$ is the relativistic degrees of freedom, and we use 
\begin{equation}
 \begin{split}
   s = \frac{2 \pi^2}{45} g_*(T) T^3 ~,~ H = 1.66 g_*^{1/2} \frac{T^2}{m_{pl}} . 
 \end{split}
\end{equation}
We obtain the relic density of stau at the BBN era by integrating these equations 
from $T_{f}$ to the temperature for beginning the BBN under the initial 
condition of the total 
number density of stau and neutralino. These equations make it clear that if the 
tau number density is out of the equilibrium, the ratio between those of stau 
and neutralino does not satisfy the Eq.(\ref{ratio}).

\section{Numerical results}   \label{numerical} 

In this section, we will first show the evolution of the stau number density, and
then study the relation between the relic density of stau and the modification
of nucleosynthesis.
Finally, we study a solution of the $^7\mathrm{Li}$ problem with long-lived
stau in the coannihilation scenario based on
Ref.~\cite{Jittoh:2007fr,Bird:2007ge,Jittoh:2008eq}.

\subsection{Total abundance}   \label{total abundance} 

\begin{figure} 
\begin{center}
\includegraphics[width=250pt, clip]{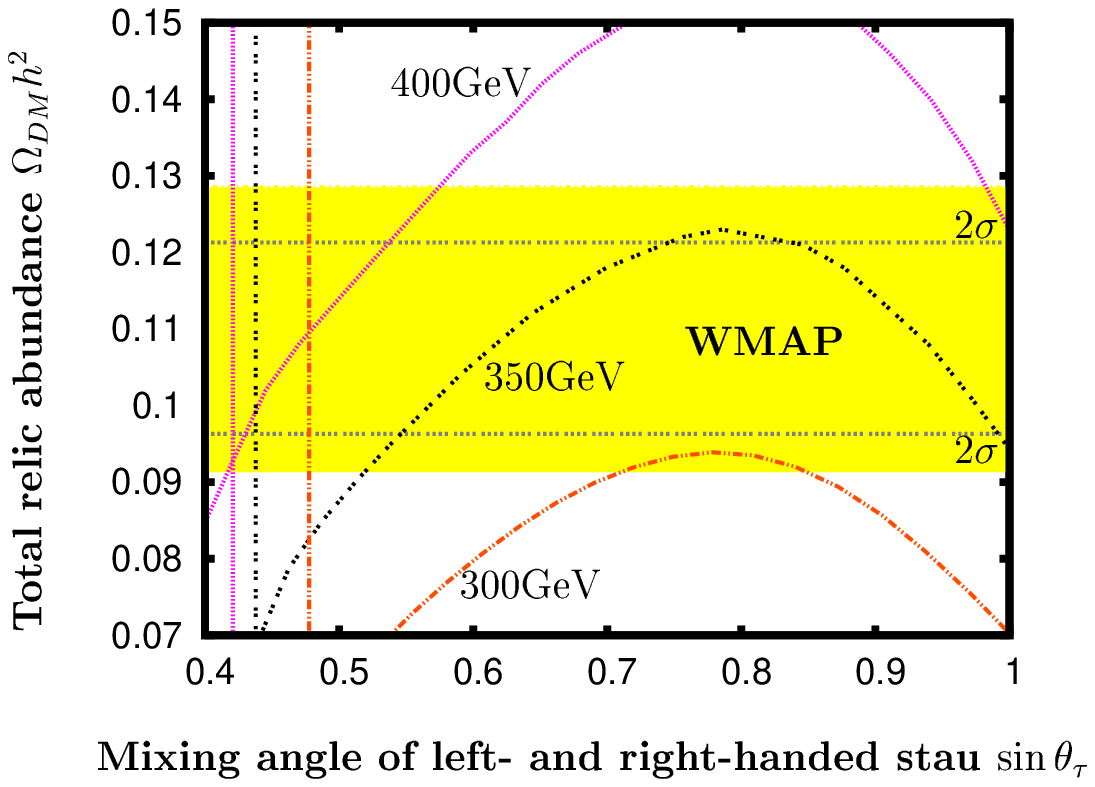}
\caption{{\small{%
      Total abundance of staus and neutralinos, which corresponds to the relic
      abundance of DM.  Each line shows the total abundance, and
      attached value represents the stau mass. Yellow band represents the
      allowed region from the WMAP observation at the $3\sigma$ level
      , and the region inside the horizontal dotted lines corresponds 
      to allowed region at the $2\sigma$ level \cite{Dunkley:2008ie}.
      In the left side of vertical lines, the LSP is left-handed sneutrino.
      Three lines correspond to $m_{\tilde \tau} = 400, 350, 300 \, \mathrm{GeV}$
      from left to right.
}}  }
\label{total}
\end{center}
\end{figure} 

As a first step for the calculation of the relic number density of stau, based on the
discussion in section \ref{evo}, we calculate the total abundance of stau and
neutralino with micrOMEGAs \cite{Belanger:2008sj}.  Fig. \ref{total} shows the
total abundance, which corresponds to the relic abundance of DM, as a
function of $\sin \theta_{\tau}$, where $\theta_{\tau}$ is the mixing angle
between left and right-handed stau.  Each curved line shows the total abundance
for each stau mass, and horizontal band represents the allowed region from the
WMAP observation at the 3$\sigma$ level ($0.0913 \leq \Omega_{DM} h^2 \leq
0.1285$), and the region inside the horizontal dotted lines corresponds 
to allowed region at the $2\sigma$ level  ($0.0963 \leq \Omega_{DM} h^2 \leq
0.1213$) \cite{Dunkley:2008ie}.
In the left side of vertical lines, the LSP is the left-handed sneutrino.
Three lines correspond to $m_{\tilde \tau} = 400, 350, 300 \,\mathrm{GeV}$
from left to right.
Since the left-handed sneutrino DM has been ruled out by constraints from the
direct detection experiments \cite{Falk:1994es}, we focus on the right-side
region. Here we took $\gamma_\tau = 0$ and $\delta m = 100 \, \mathrm{MeV}$.

The total abundance increases first as the heavier stau mixes to the lighter
stau, then turns to decrease at $\sin\theta_{\tau} \simeq 0.8$.  The increase
of the abundance can be understood by the fact that the annihilation cross section 
of $\tilde{\tau} + \tilde{\tau} \rightarrow \tau + \tau$ becomes smaller as the
heavier stau mixes. The increase is gradually compensated by two annihilation
processes, $\tilde{\tau} + {\tilde{\tau}}^\ast \rightarrow W^+ + W^-$ and 
$\tilde{\chi}^0_1 + \tilde{\tau} \rightarrow W^- + \nu_\tau$, as the left-right 
mixing becomes large. The latter process can not be ignored because left-handed 
sneutrino is 
degenerate to stau and neutralino in the present parameter set. These processes 
become significant for $\sin\theta_{\tau} < 0.8$. Another process which reduces 
the total abundance due to the left-right mixing is $\tilde{\tau} + 
\tilde{\tau}^\ast \rightarrow t + \bar{t}$ through s-channel exchange of the 
heavy Higgses. This annihilation process becomes significant as the mixing reaches 
to $\pi/4$ and the masses of two staus are split. In fig.~2, the mass difference 
between the lighter and the heavier stau is fixed to be $30$ GeV to maximize the 
DM abundance, but the same result can be obtained by changing the stau 
mass for another values of the mass difference.
As shown in Fig. \ref{total}, the total abundance also strongly depends on
$m_{\tilde \tau}$.
This is understood as follows.
In the non-relativistic limit, since the relic number density of relic species is 
proportional to $(m_{relic} \langle \sigma v \rangle_{\textrm{sum}})^{-1}$ 
and $\langle \sigma v \rangle_{\textrm{sum}}$ (Eq. (\ref{f})) is proportional 
to $1/m_{relic}^2$ \cite{Kolb}, the total number density $N$ is proportional to 
$m_{\tilde \tau}$,
\begin{equation}
\begin{split}
    N \propto \frac{1}{m_{\tilde \tau} \langle \sigma v \rangle_{\textrm{sum}}} 
    \propto \frac{1}{1/m_{\tilde \tau}}
    = m_{\tilde \tau} ~,
 \end{split}
 \label{k}
\end{equation}
and the total abundance is given by $\Omega_{DM} h^2 \sim m_{\tilde \tau} N$.
Thus the total abundance is proportinal to $m_{\tilde \tau}^2$, and it is
consistent with the result in Fig. \ref{total}.

\subsection{Stau relic density at the BBN era}   \label{stau density}  

\begin{figure} 
\begin{center}
\includegraphics[width=250pt, clip]{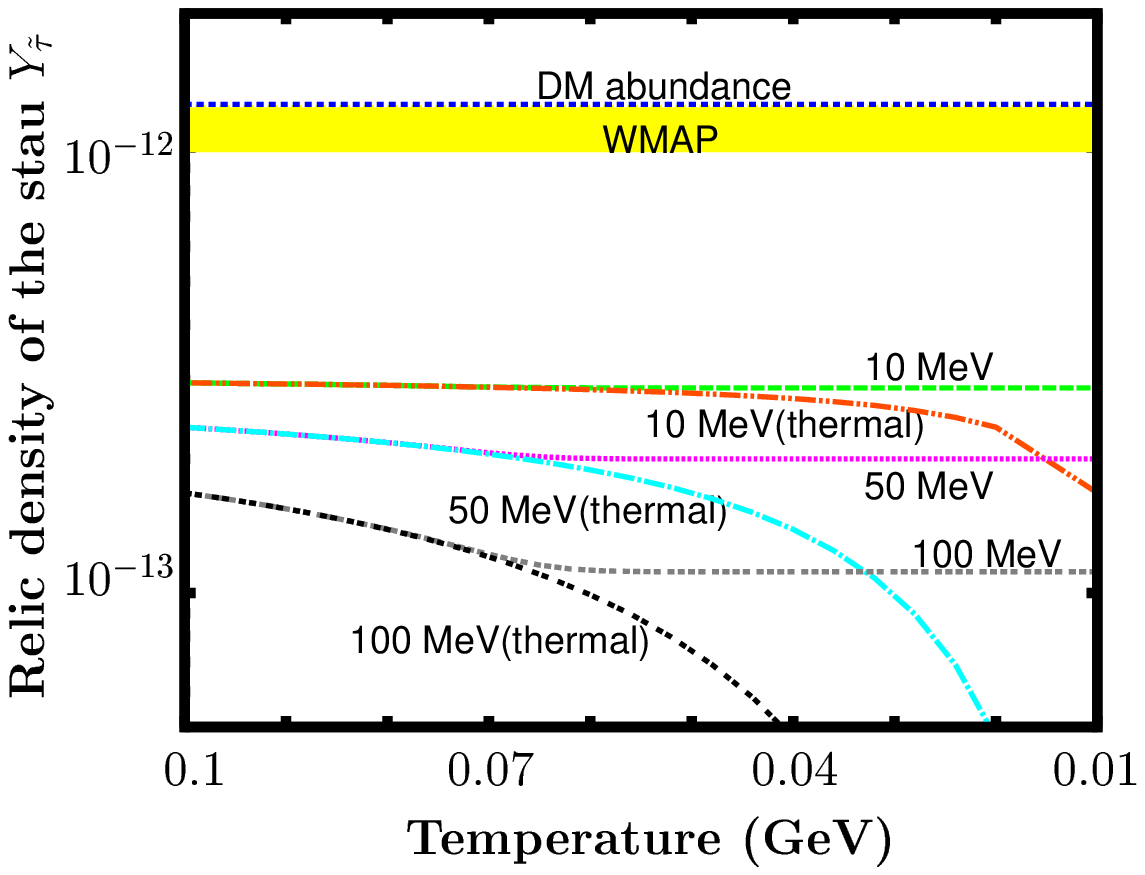}
\caption{{\small{%
      The evolution of the number density of negative charged stau.
      Each line attached [$\delta m$] shows the actual evolution of the 
      number density of stau, while the one atattched [$\delta m$(thermal)] shows its
      evolution under the equilibrium determined given by Eq.~(\ref{ratio}) and
      the total relic abundance. 
      Yellow band represents the allowed region from the WMAP observation at the
      $2\sigma$ level \cite{Dunkley:2008ie}.}} }
\label{abu}
\end{center}
\end{figure}

Next, we solve the Boltzmann equations (\ref{h}), (\ref{i}), and (\ref{j})
numerically, and obtain the ratio of the stau number density to the total number
density of stau and neutralino.
Fig. \ref{abu} shows the evolution of the number density of stau as a
function of the universe temperature. Here we took $m_{\tilde \tau} = $
350 GeV, $\sin \theta_\tau$ = 0.8, and $\gamma_\tau $ = 0 and chose
$\delta m$ = 10 MeV, 50 MeV, and 100 MeV as sample points.
Each line attached [$\delta m$] shows the actual evolution of the 
number density of stau, while the one atattched [$\delta m$(thermal)] shows its
evolution under the equilibrium determined by Eq.~(\ref{ratio}) and the
total relic abundance.
Horizontal dotted line represents the relic density of DM, which is the
total abundance calculated above. We took it as a initial condition of total
value for the calulation of the number density ratio. Yellow band represents the
allowed region from the WMAP observation at the $2\sigma$ level
\cite{Dunkley:2008ie}.

The number density evolution of stau is qualitatively understood as follows. As
shown in Fig. \ref{abu}, the freeze-out temperature of stau almost does not
depend on $\delta m$.
It is determined by the exchange processes Eq. (\ref{exchange}), whose
magnitude $\langle \sigma v \rangle Y_{\tilde \tau} Y_{\gamma}$ is governed by
the factor $e^{-(m_\tau - \delta m)/T}$, where $m_\tau$ represents the tau lepton
mass.
The freeze-out temperature of the stau density $T_{f \textrm{(ratio)}}$ is given
by $(m_{\tau} - \delta m)/T_{f \textrm{(ratio)}} \simeq 25$ as in Eq.~(\ref{g}),
since the cross section of the exchange process is of the same magnitude as weak
processes.
Thus $T_{f \textrm{(ratio)}}$ hardly depends on $\delta m$.
In contrast, the ratio of the number density between stau and neutralino depends
on $\delta m$ according to Eq.~(\ref{ratio}), $n_{\tilde \tau} / n_{\tilde \chi}
\sim \text{exp} (- \delta m / T)$, since they follow the Boltzmann distribution
before their freeze-out.
Thus, the relic density of stau strongly depends on $\delta m$.

Here, we comment on the dependence of the stau relic density $n_{\tilde \tau^-}$
on other parameters such as $m_{\tilde \tau}$, $\theta_{\tau}$, and
$\gamma_{\tau}$. 
The number density of the negatively charged stau is expressed in terms of the total
relic density $N$ by
\begin{equation}
 \begin{split}
   n_{\tilde \tau^-} 
   =
   \frac{N}{2 (1 + e^{\delta m/T_{f \textrm{(ratio)}}})} ~.
 \end{split}     \label{l}
\end{equation} 
Here, the freeze-out temperature $T_{f \textrm{(ratio)}}$ hardly depends on
these pararameters.
This is because the cross section of the exchange processes are changed by these
parameters at most by factors but not by orders of magnitudes, and the $T_{f
  \textrm{(ratio)}}$ depends logarithmically on $\langle \sigma v \rangle$ as
shown in Eq.~(\ref{g}).
On the other hand, the total relic density $N$ is proportional to $m_{\tilde
  \tau}$ as in Eq.~(\ref{k}).
The value of $N$ is also affected by the left-right mixing $\theta_{\tau}$ as
seen in Fig.~\ref{total} since the annihilation cross section depends on this
parameter.
In contrast, $\gamma_{\tau}$ scarcely affects the relic density, since this
parameter appears in the annihilation section through the cross terms of the
contributions from the left-handed stau and the right-handed one, and such terms
always accompany the suppression factor of $m_{\tau}/m_{\tilde{\tau}}$
compared to the leading contribution.
Thus the relic number density of stau $n_{\tilde{\tau}}$ strongly depends on
$m_{\tilde{\tau}}$ and $\theta_{\tau}$ while scarecely depends on
$\gamma_{\tau}$.

We comment on the generality of our method to calculate the density of exotic
heavy particles that coannihilate with other (quasi)stable particles: we
calculate the total number density of these particles and then calculate
the ratio among them by evaluating the exchange processes such as
Eq.~(\ref{exchange}).
This method of calculation can be found versatile in various scenarios including
the catalyzed BBN and the exotic cosmological structure formation
~\cite{Sigurdson:2003vy,Profumo:2004qt,Kohri:2009mi}.

\subsection{Long-lived stau and BBN}   \label{stau BBN}  

After the number density of stau freezes out, stau decays according to its
lifetime \cite{Jittoh:2005pq}, or forms a bound state with a nuclei in the BBN
era.
Their formation rate has been studied in literatures~\cite{Kohri:2006cn,
  Jittoh:2007fr, Bird:2007ge}.
The bound states modify the predictions of SBBN, and make it possible to solve
the $^7$Li problem via internal conversion processes in the bound state
\cite{Jittoh:2007fr, Bird:2007ge, Jittoh:2008eq}.

\begin{figure} [t!]
\begin{center}
\includegraphics[width=235pt,clip]{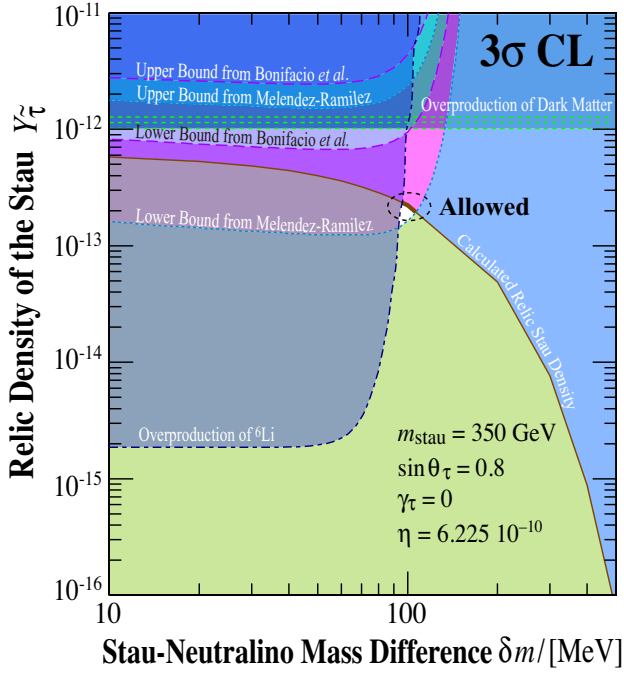}
\caption{ %
  Parameter regions that are consistent with the observed abundances of
  the DM and the light elements.
  Constraints from the observed $^7$Li abundances are due to Bonifacio
  \textit{et al.}~\cite{Bonifacio:2006au} and Melendez and
  Ramilez~\cite{Melendez:2004ni}.
  The calculated relic number density of stau with the indicated parameter
  is also shown. 
  The top left region is excluded from $^{6}$Li overproduction for
  $Y_{\tilde{\tau}^{-}} \gtrsim 10^{-15}$ and $\delta m \lesssim$ 100~MeV.
}
\label{fig:allowed3sigma}
\end{center}
\end{figure}

In Fig.~\ref{fig:allowed3sigma} we show parameter regions that are consistent
with the observed abundances of the DM and of the light elements.
We calculate the relic density of stau by varying the value of $\delta m$
with the values of $m_{\tilde \tau} = $ 350 GeV, $\sin \theta_\tau$ = 0.8, and
$\gamma_\tau$ = 0.
With these parameters, the allowed region is shown inside the dotted oval. 
We see that there are allowed regions at $Y_{\tilde{\tau}^{-}} \sim
10^{-13}$ -- $10^{-12}$ for $\delta m \lesssim$~130~MeV to solve the
$^7$Li problem at 3$\sigma$.
On the other hand, it is found that the observational $^6$Li to $^7$Li ratio
excludes $Y_{\tilde{\tau}^{-}} \gtrsim 10^{-15}$ and $\delta m \lesssim$
100~MeV.
We will explain this feature as follows.

We have adopted following observatinal abundances of  $^6$Li and
$^7$Li.
Throughout this subsection, $n_{i}$ denotes the number density of a particle
``$i$'', and observational errors are given at $1 \sigma$.
For the $n_{\rm ^6Li}$ to $n_{\rm ^7Li}$ ratio, we use the upper
bound~\cite{Asplund:2005yt},
\begin{eqnarray}
    \label{eq:Li6obs}
    \left( n_{\rm ^6Li}/n_{\rm ^{7}Li} \right)_p
     < 0.046 \pm 0.022 + 0.106,
\end{eqnarray}
with a conservative systematic error (+0.106)~\cite{Hisano:2009rc}.
For the $^7$Li abundance, we adopt two observational values of the $n_{\rm
  ^7Li}$ to $n_{\rm H}$ ratio.
Recently it has been reported to be
\begin{eqnarray}
    \label{eq:Li7obs}
    \log_{10}(n_{^{7}{\rm Li}}/n_{\rm H})_{p}=
    -9.90 \pm 0.09,
\end{eqnarray}
by Ref.~\cite{Bonifacio:2006au}, and on the other hand, a milder one was
also given by Ref.~\cite{Melendez:2004ni},
\begin{eqnarray}
    \label{eq:Li7obs_MR}
    \log_{10}(n_{^{7}{\rm Li}}/n_{\rm H})_{p}=
    -9.63 \pm 0.06.
\end{eqnarray}

In the current scenario $^6$Li can be overproduced by the scattering
of the bound state $(^{4}{\rm He}\tilde{\tau}^{-})$ off the background
deuterium through $(^{4}{\rm He}\tilde{\tau}^{-}) + {\rm D} 
\rightarrow ^6{\rm Li} + \tilde{\tau}^{-}$~\cite{Pospelov:2006sc}.
The abundance of the nonthermally-produced $^{6}$Li through this process is
approximately represented by
\begin{eqnarray}
    \label{eq:DeltaYLi6}
    \Delta Y_{^{6}{\rm Li}} \sim \frac{\langle \sigma v
    \rangle_{^{6}{\rm Li}}n_{\rm
    D}}{H} Y_{\tilde{\tau}^{-}},
\end{eqnarray}
with $\langle \sigma v \rangle_{^{6}{\rm Li}}$ the thermal average of the cross
section times the relative velocity for this process~\cite{Hamaguchi:2007mp},
and $n_{\rm D}$ the number density of deuterium.
By using (\ref{eq:Li6obs}), we see that the additional $^6$Li production is
constrained to be $\Delta Y_{^{6}{\rm Li}} < {\cal O}(10^{-21})$.
Numerical value of $\langle \sigma v \rangle_{^{6}{\rm Li}}$ gives $\langle
\sigma v \rangle_{^{6}{\rm Li}} n_{\rm D}/ H \sim {\cal O}(10^{-6})$ at $T\sim
$10 keV.
Then from (\ref{eq:DeltaYLi6}) it is found that the upper bound on the abundance
of stau should be $Y_{\tilde{\tau}^{-}} \lesssim 10^{-15}$.
Because the bound state $(^{4}{\rm He}\tilde{\tau}^{-})$ forms at $T \lesssim
$10 keV, this process is strongly constrained for $\tau_{\tilde{\tau}^{-}}
\gtrsim 10^{4}$ with $\tau_{\tilde{\tau}^{-}}$ being the stau lifetime, which
corresponds to $\delta m \lesssim 100$ MeV.
Note that the ratio $\langle \sigma v \rangle_{^{6}{\rm Li}} n_{\rm D}/ H $
rapidly decreases as the cosmic temperature decreases, and this nonthermal
production of $^{6}$Li is much more effective just after formation of the bound state.
This is the reason why we can estimate (\ref{eq:DeltaYLi6}) at around
10 keV.

On the other hand, the rates of $^{7}$Be and $^{7}$Li destruction through the
internal conversion~\cite{Jittoh:2007fr,Bird:2007ge,Jittoh:2008eq} could
be nearly equal to the formation rates of the bound sate $(^{7}{\rm
  Be}\tilde{\tau}^{-})$ and $(^{7}{\rm Li}\tilde{\tau}^{-})$, respectively.
This is because the timescale of the destruction through the internal conversion
is much faster than that of any other nuclear reaction rates and the Hubble
expansion rate.
Then the destroyed amount of $^{7}$Be (or $^{7}$Li after its electron capture)
is approximately represented by
\begin{eqnarray}
    \label{eq:Delta7}
   \Delta Y_{^{7}{\rm Be}} \sim  \frac{\langle \sigma v \rangle_{\rm
    bnd,7}n_{^{7}{\rm Be}}}{H} Y_{\tilde{\tau}^{-}},
\end{eqnarray}
where $\langle \sigma v \rangle_{\rm bnd,7} \sim 10^{-2} {\rm
GeV}^{-2} (T/30{\rm keV})^{-1/2}(Z/4)^{2} \times(A/7)^{-3/2}
(E_{b^{7}{\rm Be}}/1350 {\rm keV})$ is the thermally-averaged cross
section times the relative velocity of the bound-state formation for
$(^{7}{\rm Be}\tilde{\tau}^{-})$ ~\cite{Kohri:2006cn,Bird:2007ge}. We
request $\Delta Y_{^{7}{\rm Be}}$ to become $\sim {\cal O}(10^{-20})$
to reduce the abundance of $^{7}{\rm Be}$ to fit the observational
data. Then the abundance of $\tilde{\tau}^{-}$ should be the order of
$ \Delta Y_{^{7}{\rm Be}} (\langle \sigma v \rangle_{\rm bnd,7}
n_{^{7}{\rm Be}}/H )^{-1} \sim {\cal O}(10^{-12})$ with $\langle
\sigma v \rangle_{\rm bnd,7} n_{^{7}{\rm Be}}/H \sim 10^{-8}$ at $T$ =
30 keV. Because $\langle \sigma v \rangle_{\rm bnd,7} n_{^{7}{\rm
Be}}/H$ decreases as the cosmic temperature decreases ($\propto
T^{1/2}$), the destruction is more effective just after the formation of
The bound state. This validates that we have estimated
(\ref{eq:Delta7}) at 30 keV. Therefore the parameter region at around
$Y_{\tilde{\tau}^{-}} \sim 10^{-12}$ and $\delta m \lesssim $~130~MeV
is allowed by the observational $^{7}$Li. Here $\delta m \lesssim
$~130~MeV corresponds to $\tau_{\tilde{\tau}} \gtrsim 10^{3}$~s. The
case for the destruction of ($^{7}$Li$\tilde{\tau}^{-}$) through the
internal conversion is also similar to that of
($^{7}$Be$\tilde{\tau}^{-}$)~\cite{Jittoh:2007fr,Jittoh:2008eq}.

Further constraints come from the relic density of the DM, which can be stated in
terms of the stau relic density.
It is calculated as shown in Fig.~\ref{fig:allowed3sigma} for the present values
of parameters.
Applying all the constraints, we are led to the allowed interval shown by the
thick line in the figure.

\subsection{Constraint on parameter space of stau}
\label{LHC} 

\begin{figure} [t!]
\begin{center}
\includegraphics[width=250pt,clip]{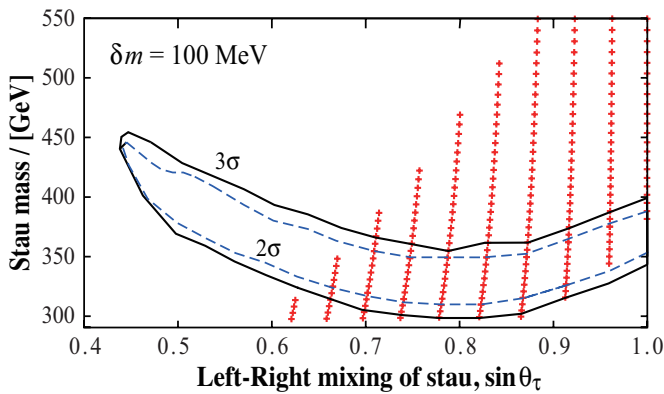}
\caption{ %
  Parameter space in which the
  calculated abundances of the DM and that of the light elements are consistent
  with their values from the observations.
  Parameter region surrounded by black solid (blue dashed) line shows
  the allowed region from the WMAP observation at the 3$\sigma$
  (2$\sigma$) level \cite{Dunkley:2008ie}.  Red crisscross points show
  the parameters which is consistent with observational abundances for the
  light elements including $^7$Li at $3\sigma$ level. } \label{allowed}
\end{center}
\end{figure}

Finally, we show in Fig.~\ref{allowed} the parameter space in which the
calculated abundances of the DM and that of the light elements are consistent
with their values from the observations.
Here, based on the discussion in previous subsection, we took $\delta m =
100 \,\mathrm{MeV}$.
Parameter region surrounded by black solid (blue dashed) line is allowed
by the relic abundance of the DM from the WMAP observation at the $3\sigma$
($2\sigma$) level \cite{Dunkley:2008ie}, which corresponds to $0.0913
\leq \Omega_{\textrm{DM}} h^2 \leq 0.1285$ ($0.0963 \leq
\Omega_{\textrm{DM}} h^2 \leq 0.1213$).
Red crisscross points show the parameters which are consistent with the observational
abundances for the light elements including $\mathrm{^7Li}$, where the observational
$\mathrm{^7Li}$ abundance is yielded by literature \cite{Melendez:2004ni}.

The abundance of the light elements constrains the parameter space due to the
following reasons.
First, the region where the stau mass is less than 300 GeV is excluded since the
relic density becomes too small to destruct $\mathrm{^{7}Li}$ sufficiently. 
Next, the top-left region of the figure is excluded since the lifetime of
stau becomes too long and hence overproduces $\mathrm{^6Li}$ through the
catalyzed fusion \cite{Pospelov:2006sc}. 
Indeed, the lifetime of stau gets longer as its mass gets heavier due to
the small phase space of the final state \cite{Jittoh:2005pq}.
On the other hand, its lifetime gets shorter as the left-right mixing angle
increases in the present parameter space \cite{Jittoh:2005pq}.
As a result, the final allowed region becomes as shown by the red crisscrosses
in Fig.~\ref{allowed}.

The black solid curve and the blue dashed curve are the constraints from the relic
abundance of the DM as discussed in subsection \ref{total abundance}.
Note that the relic abundance is insensitive to the mass difference $\delta m$ for 
$\delta m << m_{\tilde \chi}$.
Combination of the constraints on the abundance of the light elements
and of the DM strongly restricts the allowed region and leads to
$\delta m \simeq 100 \,\mathrm{MeV}$, $m_{\tilde \tau} = (300 \textrm{
-- } 400) \,\mathrm{GeV}$, and $\sin\theta_{\tau} = (0.65 \textrm{ -- }
1)$.
In Fig.~\ref{fig:allowed3sigma}, these parameter values correspond to
the white triangular region below the allowed region (thick line).
Our model can thus provide a handle to the mixing angle, which has few
experimental signals, once the value of $m_{\tilde \tau}$ is determined.

\section{Summary and discussion}   \label{summary} 

We have studied the evolution of stau number density in the MSSM 
coannihilation scenario, in which the LSP and the NLSP are the lightest 
neutralino and the lighter stau, respectively, and have a small mass 
difference $\delta m \lesssim \mathcal{O}$(1GeV).  In this case, 
stau can survive until the BBN era, and provide additional 
nucleosynthesis processes.  It is therefore necessary to see how large 
the relic density of stau is at the BBN era.

We have shown the Boltzmann equations for the calculation of the relic number
density of stau, and have found that the number density of stau continues to evolve
through the exchange processes Eq.(\ref{exchange}), even after the relic
abundance of the DM is frozen out. Thus, we need to calculate the stau relic
density by a two-step procedure. In the first step, we calculate the
total abundance of stau and neutralino, which corresponds to the relic
abundance of the DM. The total abundance is controlled only by pair
annihilation processes of the supersymmetric particles.  In the second step,
we calculate the ratio of the stau number density to the total number
density of stau and neutralino, which is governed only by the exchange
processes. We have calculated the relic density of stau at the BBN era
by solving the Boltzmann equations numerically.
The freeze-out temperature $T_{f \textrm{(ratio)}}$ is determined by 
$(m_{\tau} - \delta m)/T_{f \textrm{(ratio)}} \simeq 25$ and the 
relic density of stau are given by Eq.~(\ref{l}). Thus it becomes larger 
as the mass difference between stau and neutralino gets smaller.
Our method of calculation is generally applicable to obtain the density
of exotic heavy particles that coannihilate with other (quasi)stable
particles: we calculate the total number density of these particles and
then calculate the ratio among them by evaluating the exchange processes
such as Eq.~(\ref{exchange}).
This method of calculation can be found versatile in various scenarios
including the catalyzed BBN and the exotic cosmological structure
formation.

At the BBN era, the long-lived stau form bound states with nuclei, and 
provide exotic nucleosynthesis processes. One of them is the internal 
conversion process, which offers a possible solution to the 
$\mathrm{^7Li}$ problem. Applying the calculated relic density of stau, 
we have calculated the primordial abundance of light elements 
including these exotic processes. We have found the parameter space 
consistent with both of the calculational results and the observations for 
the relic abundance of the DM and the light elements abundance including 
$\mathrm{^7Li}$. 
We have shown a prediction for the values of the parameters relevant to
stau and neutralino, which is shown in Fig. \ref{allowed}.  Consistency
between the theoretical prediction and the observational result, both of the DM
abundance and the light elements abundance requires $\delta m \simeq 100
\,\mathrm{MeV}$, $m_{\tilde \tau} = (300 \textrm{ -- } 400) \,
\mathrm{GeV}$, and $\sin\theta_{\tau} = (0.65 \textrm{ -- } 1)$.

\section*{Acknowledgments}   

The work of K. K. was supported in part by PPARC Grant
No. PP/D000394/1, EU Grant No. MRTN-CT-2006-035863,  the European
Union through the Marie Curie Research and Training Network
``UniverseNet,'' MRTN-CT-2006-035863, and Grant-in-Aid for Scientific
research from the Ministry of Education, Science, Sports, and Culture,
Japan, No. 18071001.
The work of J. S. was supported in part by the Grant-in-Aid for the Ministry of Education, Culture, Sports, Science, 
and Technology, Government of Japan (No. 20025001, 20039001, and 20540251). 
The work of T. S. was supported in part by MEC and FEDER (EC) Grants No. FPA2005-01678.
The work of M. Y. was supported in part by the Grant-in-Aid for the Ministry of Education, Culture, Sports, Science, 
and Technology, Government of Japan (No. 20007555).


\end{document}